\documentstyle[12pt]{article}
\newcommand{\be}{\begin{equation}}
\newcommand{\ee}{\end{equation}}
\newcommand{\ba}{\begin{eqnarray}}
\newcommand{\ea}{\end{eqnarray}}

\begin{document}
\begin{flushright}
Preprint SPbU-IP-94-03
\end{flushright}

\vspace{1.5cm}

\begin{center}
{\Large\bf Second Order Derivative Supersymmetry\\
 and Scattering Problem}\\

\vspace{1cm}

{\bf A. A. Andrianov\footnote{Department of Theoretical
Physics, University of Sankt-Petersburg,198904 Sankt-Petersburg, Russia.
E-mail: andriano@onti.phys.lgu.spb.su;\hspace{3ex}
ioffe@onti.phys.lgu.spb.su},
F. Cannata\footnote{Dipartimento di Fisica and INFN, Via Irnerio 46,
40126 Bologna, Italy. E-mail: cannata@bo.infn.it},
J.-P. Dedonder\footnote{Laboratoire de Physique Nucl\'eaire,
Universit\'e Paris 7,\, 2 place Jussieu, F - 75251 Paris Cedex 05, France
and Division de Physique Th\'eorique, Institut de Physique Nucl\'eaire,
F 91406 Orsay,France.
E-mail: dedonder@ipnvax.in2p3.fr}\\
       and\\
       M. V. Ioffe$^1$}
\end{center}

\vspace{1.cm}

\begin{abstract}
Extensions of standard one-dimensional supersymmetric quantum
mechanics are discussed. Supercharges involving higher order
derivatives are introduced leading to an algebra which incorporates
a higher order polynomial in the Hamiltonian. We study scattering
amplitudes for that problem. We also study the role of a dilatation
of the spatial coordinate leading to a $q$-deformed supersymmetric
algebra. An explicit model for the scattering amplitude is
constructed in terms of a hypergeometric function which
corresponds to a reflectionless potential with infinitely
many bound states.
\end{abstract}
\newpage

\vspace{1.cm}

\section{Introduction}
\vspace{.5cm}
\hspace*{3ex}The search for dynamical connections between different
quantum systems (or subsystems) is one of the most interesting and
fundamental problems in modern quantum mechanics. Such connections are
essential to establish in order to investigate the spectral properties of
certain quantum models as well as to generate new systems with given
spectral characteristics.

     The first step in developing explicit relations between different
spectral problems was realized in the remarkable papers by Moutard
  and by Darboux \cite{darboux}
on the spectral properties of the Sturm-Liouville differential problem,
more than hundred years ago. The corresponding approach in non
relativistic quantum theory  was introduced in the forties by Schr\"odinger
\cite{schrod} for the one dimensional harmonic
oscillator problem and is known as the factorization method. It was
extensively developed by Infeld and Hull \cite{infeld} who classified
one-dimensional solvable problems.

      Another useful look at the spectral equivalence of Hamiltonians
intertwined by Darboux transformations originated from the supersymmetrical
approach to the above problem \cite{witten}, \cite{sal}. The supersymmetric
quantum mechanics (SSQM) as such has been
introduced by Witten \cite{witten} in 1981 as a toy model to illustrate
the problem of supersymmetry breaking in the general framework of
supersymmetric quantum field theories. In fact, the supersymmetric quantum
mechanics
allows to combine two isospectral Hamiltonians, which differ at most
by a single bound state, into a single Schr\"odinger equation at the
price of introducing additional fermionic degrees of freedom. The two
supersymmetric partners are related by a particular Darboux
transformation. This last approach appears to be fruitful in
providing possibilities of generalizing to multidimensional problems the
one dimensional Darboux transformations or the traditional factorization
method \cite{abi}. The operator approach based on the factorization method
allows to unravel the detailed structure of supercharges,
superhamiltonians and Hilbert superspaces.
SSQM has now developed in its own right as a source for establishing
relations between the wave functions and various observables of different
physical systems with identical energy spectra but for a few states
(see for example \cite {pursey}, \cite{genref}). There are actually
many different ways \cite{pursey}
of constructing families of isospectral Hamiltonians, generated for
instance by
iterative applications of basic procedures like the Darboux one
\cite{crum}, or such as
the Abraham and Moses one based upon the Gel'fand-Levitan equation and the
Pursey one, to quote some.

Recently extensions of SSQM have been elaborated using different
realizations of the intertwining (super)generators. In the first one
 \cite{ais}, supercharges are constructed in terms of higher-derivative
operators (HSSQM) and the corresponding superalgebra (HSUSY) becomes
polynomial in the Hamiltonian. In the second extension a dilatation
of the coordinates is involved in building supercharges \cite{spir}.
The corresponding algebra obeys $q$-deformed
SUSY relations. Both extensions open new ways for exploring quantum
mechanical systems with related energy spectra and wave functions.

The aim of our paper is to analyze links between these new approaches
and to examine their features in the scattering regime. Attention is
paid to the Witten criterion of spontaneous SUSY breaking which is
no longer applicable to the case of HSUSY. In Section 2 we review
the basics of one-dimensional SSQM. The Witten
criterion is formulated in its conventional form. In Section 3 SSQM
with supercharges of second-order in derivatives is built in the most
general form and it is found that there exist cases where the
superpartners cannot be constructed by iterations of two ordinary
Darboux transformations. Thus we find new irreducible elements, not
discussed in \cite{ais}, for building the HSUSY systems with polynomial
superalgebras. We also discuss within the one dimensional problem the
radial case. In Section 4 the consequences for scattering properties
of the partner systems in one dimensional SUSY and HSUSY are described.
In the radial problem, the necessary restrictions
which provide the correct properties of potentials and wave
functions are investigated. The pecularities of the Witten index
in second-derivative SSQM are discussed. In Section 5 the
$q$-deformation of SUSY and HSUSY induced by the dilatation of the
coordinate is considered. The interrelation between $q$-deformed
SUSY algebra with the true Hamiltonian and the ordinary
SUSY algebra with $q$-deformed Hamiltonian is established
and further generalized onto the higher-derivative SUSY.
The relation between scattering amplitudes of two $q$-superpartners is
displayed as well. In conclusion, the condition of self-similarity for
potentials is briefly summarized and the dual condition of
self-similarity for scattering amplitudes is introduced. It is solved, in
the reflectionless case, within  $q$-deformed SSQM, in terms of a
hypergeometric function.

\section{SUSY Quantum Mechanics in one dimension}
\hspace*{3ex}SSQM is generated \cite{witten},\cite{sal} by supercharge
operators $Q^+$
and $Q^- = (Q^+)^{\dag}$ which together with the Hamiltonian $H$ of
the system fulfil the algebra
\ba
(Q^{\pm})^2 = 0,
\hspace{1cm}
[ H, Q^{\pm}] = 0, \label{basic1}\\
\{Q^+, Q^-\} = H = Q^2 \label{basic2}
\ea
where we have introduced the hermitian supercharge operator, $Q = Q^+ + Q^-$.

The one-dimensional representation is realized by the
$2 \times 2$ supercharges
\be
Q^{-} = \left( \begin{array}{cc}
0 & 0 \\ a^- & 0 \end{array} \right)
\hspace{1cm}\mbox{and}\hspace{1cm}
Q^{+} = \left( \begin{array}{cc}
0 & a^{+} \\ 0 & 0 \end{array} \right)
\ee
where
\be
a^{\pm} = {\pm} \partial + W(x) \label{superp}
\ee
and the superhamiltonian is assembled from two ordinary Schr\"odinger
Hamiltonians,
\ba
H & =& \left(\begin{array}{cc}
h_{1} & 0 \\ 0 & h_{2} \end{array}\right) =
\left(\begin{array}{cc}
a^+ a^- & 0 \\ 0 & a^- a^+ \end{array}\right) =
- {\partial}^2 + \left(\begin{array}{cc}
W^2 + W'& 0 \\ 0 & W^2 - W' \end{array}\right)\nonumber\\
\nonumber\\
& =& (- \partial^2 + W^2) {\bf 1} + \sigma_3 \, W^{\prime} \label{h1h2}\\
\nonumber\\
h_i &\equiv& - \partial^2 + V_i (x), \nonumber
\ea
where $\sigma_3$ is a Pauli matrix.

A direct consequence of Eqs. (\ref{basic1}),(\ref{basic2})
is that all eigenvalues of $H$
are non-negative. In terms of components, the algebra
defined by Eqs. (\ref{basic1}),(\ref{basic2}) means that
the Hamiltonians $h_1$ and $h_2$ in Eqs. (\ref{h1h2})
are factorized. To express it differently, we can say that
to a given factorizable Hamiltonian $h_1$, one can associate a
supersymmetric partner $h_2$ such that both partners are linked by the
intertwining relations
\footnote{The factorization that follows from Eqs. (\ref{h1h2}) defines
the Hamiltonians up to an overall constant which is related to the
arbitrariness of the energy scale. We have chosen here
this constant to be zero. }
\be
h_1 a^+ = a^+ h_2 \hspace{1cm}\mbox{and}\hspace{1cm}
a^- h_1 = h_2 a^-. \label{intertw}
\ee
These relations lead to the double degeneracy of all
positive energy levels of $H$
belonging to the "bosonic" or "fermionic" sectors specified
by the grading operator $\tau = (-)^{\widehat{N_F}} = \sigma_3$
where $\widehat{N_F}$ is the fermion number operator. The grading
operator commutes with the
Hamiltonian $H$ and anticommutes with the supercharge $Q$. Thus the
supercharge operator transforms eigenstates with $\tau = + 1$ (bosons)
into eigenstates with $\tau = - 1$ (fermions) and vice versa.
Boson and fermion wave functions are eigenfunctions of $h_1$
and $h_2$ respectively. They are connected via Eqs. (\ref{intertw})
by the operators $a^{\pm}$
\be
\sqrt{E} \Psi_E^{(2)} =  a^- \Psi_E^{(1)}
\hspace{1cm}\mbox{and}\hspace{1cm}
\sqrt{E} \Psi_E^{(1)} =  a^+ \Psi_E^{(2)}. \label{psi}
\ee
The existence of zero energy states depends \cite{witten},\cite{sal} on
the asymptotics  of the superpotential $W(x)$ which appears in
Eq.(\ref{superp}). For appropriate $W(x)$, that is for particular topologies,
they can arise either in the bosonic sector, i.e.,
$$ a^- \Psi^{(1)} (x) = 0,$$
or in the fermionic one, i.e., $$ a^+ \Psi^{(2)}(x) = 0.$$
The explicit form of the solutions for the wave functions reads
\be
\Psi^{(1,2)}(x) = exp (\pm \int^x dy \ W(y))
\ee
and shows that depending on the asymptotics of the superpotential there
may be three different types of vacuum characterized by the number
of bosons $N_B$ and of fermions $N_F$ :$$N_B  = 0, N_F = 1 \, \mbox{ or}\,
N_B = 1,N_F = 0$$ for which supersymmetry is an exact symmetry and the
vacuum wave
functions are non degenerate. The third case, where the equations
$a^{\mp} \Psi^{(1),(2)} = 0$ have no normalizable solutions,
corresponds to $$N_B = N_F = 0.$$
Supersymmetry is then spontaneously broken and the vacuum states which
have positive energies are degenerate.
The above difference between unbroken and spontaneously broken supersymmetry
can be formulated in terms of an order parameter, the Witten index
\cite{witten}, which may be defined by
\be
\Delta_W = {\rm dim} \ ker \ a^+ - {\rm dim} \ ker
\ a^- = ind \ a^+ = N_B - N_F \label{deltaw}
\ee
and takes the values $0$ and $\pm 1$. Obviously, in the case of
ordinary SSQM, $\Delta_W = 0$ unambiguously characterizes the models with
spontaneously broken supersymmetry.

\section{Higher-derivative SSQM}
\subsection{One-dimensional problem}
\hspace*{3ex} We now explore other realizations where we keep
relations (\ref{basic1}) but allow for modifications of
the relation (\ref{basic2}). An example of a non-standard
realization has been introduced in \cite{ais}; it relies on the use
of higher order derivative operators in the definition of the
supercharges. Instead of the linear operators of Eq. (\ref{superp}),
let us define the second order differential operators
\be
A^+  = (A^-)^{\dagger}
= \partial^2 - 2 f(x) \partial + b(x). \label{gena+}
\ee
Then, with $Q = Q^+ + Q^-$, Eq. (\ref{basic2}) is transformed to
$$\{Q^+, Q^-\} = Q^2 = K.$$
The quasihamiltonian $K$ is thus given by the conventional
superalgebra but it is now a fourth
order differential operator, hence not of the Schr\"odinger form.

Let us assume that there exists a diagonal Hamiltonian $H$ of
Schr\"odinger type
\be
H = \left(\begin{array}{cc}
h^{(1)} & 0 \\ 0 & h^{(2)} \end{array}\right),
\ee
which commutes with supercharges $Q^{\pm}$ constructed from
$A^{\pm}$ given in (\ref{gena+}). Then it follows, from
intertwining relations similar to (\ref{intertw}),
\be
h^{(1)} A^+ = A^+ h^{(2)} \hspace{1cm}\mbox{and}\hspace{1cm}
A^- h^{(1)} = h^{(2)} A^-, \label{HA1A2}\\
\ee
that the quasihamiltonian
$K$ commutes with $H$ and is, furthermore, given in terms of $H$ by
\be
K = H^2 - 2 \alpha H + \beta \label{2SS}
\ee
where $\alpha$ and $\beta$ are constants because of the nondegeneracy of
spectra of $h^{(1)}$ and $h^{(2)}$ in the one-dimensional problem and because
$[K,Q] = 0$. The intertwining relations (\ref{HA1A2}) for $H$
require that
\be
b(x) = f(x)^2 - f'(x) -\frac{f''(x)}{2 f(x)} + \Bigl(\frac{f'(x)}
{2 f(x)}\Bigr)^2 + \frac{d}{4 f(x)^2} \label{spot}
\ee
where
\be
d = \beta -\; \alpha^2 . \label{dalf}
\ee
This condition is necessary and sufficient for the existence of a
generalized, polynomial, HSUSY algebra defined by
\be
(Q^{\pm})^2 = 0,
\hspace{0.8cm}
[ H, Q^{\pm}] = 0,
\hspace{0.8cm}
\{Q^+, Q^-\} = Q^2 = (H - \alpha)^2 + d . \label{b2}
\ee
Respectively, because of Eq. (\ref{spot}), the potentials of the superpartner
Hamiltonians are expressed solely in terms of $f(x)$ and its derivatives,
\be
V_{1,2} = \mp 2 f'(x) + f(x)^2 +
\frac{f''(x)}{2 f(x)} - \Bigl(\frac{(f'(x)}{2 f(x)}\Bigr)^2  -
\frac{d}{4f(x)^2}  + \alpha.    \label{V1,2}
\ee
The eigenfunctions of $h^{(1)}$ and $h^{(2)}$ are obtained from one
 another through the second order differential operators $A^{\pm}$ according
to Eqs. (\ref{psi}).

The existence of a relevant Schr\"odinger
operator $H$ in Eq. (\ref{2SS})
is assured, for a given quasihamiltonian $K$
\be
K = \partial^4 + \{ P(x),\partial^2\} + R(x) ,
\ee
if and only if
\be
R(x) \,-\, P^2(x) \,=\, d,
\ee
with $d$ a constant defined in (\ref{dalf}).

In particular cases  one can
factorize the elementary operators $A^{\pm}$ in terms of
ordinary superpotentials  $W_1$ and $W_2$, as in Eq. (\ref{superp}),
\be
A^+ = a_1^+ a_2^+ = (\partial + W_1(x))\, (\partial + W_2(x)).
\label{facta+}
\ee
When $\alpha = 0$ and $\beta = 0$ they are
connected by the ladder equation
\be
a_1^- a_1^+ = a_2^+ a_2^-
\hspace{1cm} \mbox{ or } \hspace{1cm}
 - W_1' + W_1^2 = W_2'
+ W_2^2. \label{ladder}
\ee

The superpotentials $W_{1,2}(x)$ are determined by
\be
W_{1,2}(x) = \pm \frac{f^{\prime} (x)}{2 f(x)} - f(x)
\ee

The factorization Eq. (\ref{facta+}) arises from two successive
standard SSQM transformations
\be
\left(\begin{array}{cc}
h^{(1)} & 0 \\ 0 & h \end{array}\right) = \left(\begin{array}{cc}
a_1^+ a_1^- & 0 \\ 0 & a_1^- a_1^+ \end{array}\right)\label{inth1}
\ee
and
\be
\left(\begin{array}{cc}
h & 0 \\ 0 & h^{(2)} \end{array}\right) = \left(\begin{array}{cc}
a_2^+ a_2^- & 0 \\ 0 & a_2^- a_2^+ \end{array}\right) \label{inth2}
\ee
together with the ladder condition (\ref{ladder}). One can then
define $H$ as
\be
H = \left(\begin{array}{cc}
a_1^+ a_1^- & 0 \\ 0 & a_2^- a_2^+ \end{array}\right).
\ee
This polynomial, higher-derivative
algebra for SSQM (HSSQM) obviously yields the
double degeneracy of the positive energy part of the spectrum of~$H$.

Nontrivial effects can be found in the lowest energy subspace of both
$h^{(1)}$ and $h^{(2)}$. In particular, in this case, the Witten criterion
(\ref{deltaw}) of spontaneous breaking of supersymmetry is not effective.
To illustrate this point, let us write the most general normalizable
solution of the zero-mode equations, i.e. $A^{\pm} \Psi_{B,F}(x) = 0$,
\be
\Psi_{B,F}(x) = A_{B,F} \sqrt{f(x)} \, exp (\pm \int_{x_0}^x dy f(y))
\ee
where $A_{B,F}$ are constants. The asymptotic behaviour of $f(x)$
determines the number of zero modes of $A^{\pm}$ and their properties.
The Witten index of operators $K$ and $H$
in the present case varies in general from $-1$ to $+1$ but the
most interesting situation arises for configurations such that
\be
f(x) \longrightarrow 0 \atop{x\rightarrow \pm\infty}
\ee
and
\be
\int_{- \infty}^{+ \infty}dy \, f(y) < \infty
\ee
One has then zero modes of operator $H$ with both fermionic and bosonic
components, $$N_B = N_F = 1$$.
We can conclude thus that the Witten criterion is not applicable to
HSSQM. Namely, this configuration has $\Delta_W = 0$ though it does
not reveal spontaneous breaking of supersymmetry due to the existence
of the doubly degenerate zero modes of $A^{\pm}$.

One can proceed to a similar analysis if one imposes a constant shift $c$
to the ladder equation
\be
a_2^+ a_2^- + c = a_1^- a_1^+  \label{shift}
\ee
where we consider $c>0$ without loss of generality.
This equation implies the following relation
\be
\{Q^+, Q^- \} = \widetilde H (\widetilde H - c) \nonumber
\ee
where to compare with Eq.(\ref {2SS})
\ba
\widetilde H = H + \frac{1}{2} (c - 2\alpha) =
\left(\begin{array}{cc}
a^+_1 a^-_1 & 0 \\ 0 & a^-_2 a^+_2 + c \end{array}\right)
\ea
The constant $c$ is actually related to $d$ defined in Eq. (\ref{dalf}) by
\be
c^2 = - 4 d \nonumber
\ee
and the shifted ladder condition is only meaningful for $d < 0$.

We consider now the case $c = 2\alpha$ which implies $\beta= 0$ \cite{ais}.
The superpotentials $W_{1,2}$ can then be parametrized in terms of a
function $f(x)$ as follows
\be
W_{1,2} = \pm\frac{2f'(x) + c}{4f(x)} - f(x). \label{W - f}
\ee
If there exist zero-modes of both supercharges $Q^{\pm}$ then they can
generally expessed in terms of $f(x)$ and $c$ as follows,
\ba
\psi_{B,F}  =  A_{B,F} \exp\biggl\lbrace \int\limits_{a}^{x}
( & \mp & f(y) + \frac{2f'(y) + c}{4f(y)} ) dy \biggr\rbrace\nonumber\\
 & + & D_{B,F}  \exp \biggl\lbrace \int\limits_{a}^{x}
(\mp f(y) + \frac{2f'(y) - c}{4f(y)} ) dy \biggr\rbrace.
\ea

The Witten index for $c \neq 0$ is defined according to Eq.(\ref{deltaw})
where $N_{B,F}$ characterize the number of zero modes of supercharges
$Q^{\pm}$ which may not be the ground states of $H$. For $c>0 (c<0)$
it can take the values between $-1$ and $+2$ ($-2$ and $+1$).

The Witten's criterion is not working again for functions $f(x)$ such
that
\ba
f(x) \longrightarrow  f_{\pm}; \quad x\rightarrow \pm\infty ; \nonumber\\
\mbox{\rm sign} f_+ = - \mbox{\rm sign} f_- = +1,
 \quad 16 f_{\pm}^4 < c^2, \label{fpm}
\ea
and
\be
f(x)\mid_{x\simeq x_l} = -\frac{1}{2} \, c (x-x_l) + o(x-x_l).
\ee
The limiting case $c=0$ is reproduced only if simultaneously
$c \rightarrow 0$ and $x_l \rightarrow \infty$.

In the case $c > 0$, the regularized Witten index \cite{witten},\cite
{cec} reads
\be
\Delta_W^{reg} = Tr [ (-)^{N_F} e^{-\gamma \widetilde H}] =
1 - e^{- \gamma c}.
\ee
and takes any values between $0$ and $1$ for positive $\gamma$.
Hence, we have seen that for HSSQM systems the Witten index does not
characterize the spontaneous breaking of supersymmetry.

As we have seen, the shifted ladder equation (\ref{shift}) is only
meaningful if the discriminant $d$ is negative; it is this property
that allows to introduce the intermediate hermitian Hamiltonian
in Eqs. (\ref{inth1}), (\ref{inth2}). If $d > 0$, no such relation as
(\ref{shift}) can be defined; we therefore refer to this class of
second order derivative SSQM as irreducible.

Such a class represents a new primitive element in building of
supersymmetrical ladders and respectively of polynomial SSQM.
Clearly, the corresponding  supercharge cannot possess any zero modes
since it is bounded from below by the constant $\sqrt{d}$, as shown by
Eqs (\ref{b2}). Therefore, for such a supersymmetric
Hamiltonian the Witten criterion is valid. In the case under
discussion the function $f(x)$ should be
assumed to be nodeless in order to avoid singular supercharges and
potentials, Eq. (\ref{V1,2}). For instance, one can make the following
ansatz, $ f(x) = \exp (x) + \exp (-2x)$ and obtain two equivalent
Hamiltonians with wave functions connected by operators $A^{\pm}$,
Eq. (\ref{gena+}).
The generalization of second order derivative SUSY algebra
to differential operators $A^{\pm}$ of higher
order is straightforward and leads to polynomials of $H$ in the right
hand side of Eq. (\ref{b2}). The ladder construction (\ref{ladder}),
(\ref{shift}) is
applicable when it is made of the primitive elements of the first and
second order in derivatives. Thus one expects, in general, the following
polynomial superalgebra,
\be
Q^2 = \prod_{i + 2j = n} (H - c_i) \bigl((H - \alpha_j)^2 + d_j
\bigr);\quad d_j > 0. \label{exten}
\ee
\subsection{Radial problem}
\hspace*{3ex}Sofar we have considered one-dimensional SSQM on the line,
\quad\mbox{$x \in (- \infty, + \infty)$}. In problems
with rotational symmetry one introduces the radial Schr\"odinger operator
on the half line, \mbox{$ r \in [0, + \infty)$}, and one can similarly
discuss features of second-derivative SSQM; for definiteness, we restrict
ourselves to the 3-dimensional case.

The radial problem differs by the boundary conditions. After reduction
of the radial part of Schr\"odinger operator
to the conventional form, Eq. (\ref{h1h2}),
one finds in standard SSQM with nonsingular potentials that in the
$l-th$ partial wave the
behavior of the wave function is regular at the origin
$\Psi \sim r^{l+1}$. Furthermore the corresponding superpotential
behaves as
$W(r) \sim (l+1)/r$ or $\sim -l/r$ for $r \sim 0$. As a
consequence the superpartners describe different partial waves:
$l_2 = l_1 \pm 1$.

Here we take a slightly more general approach allowing for different
behavior of the wave function reflecting singularities
of centrifugal type
and also possible existence of bound states at zero energy (generating
as we will see in the next section anomalies in the scattering problem).
Therefore
we only require that
the reduced radial wave function vanishes at the origin and
(for the discrete spectrum) decreases
at infinity faster than $1/\sqrt r$ .

Thus in order to construct the HSSQM Hamiltonian we assume that
\be
f(r)\mid_{r \rightarrow 0} \sim  f_0 \cdot r^{\lambda}
\ee
Acceptable values of $\lambda$ are constrained by the combination of
two requirements:\\
a) \quad the potentials $V_{1,2}$
from Eq.(\ref{V1,2}) with centrifugal terms at $ r \sim 0$
for the angular momentum $l$ and with a  core at short range,
\be
V_{1,2} \sim \mp 2 f_0 \lambda r^{\lambda - 1} + f_0^2 r^{2\lambda} +
\frac{\lambda (\lambda - 2)}{4 r^2} - \frac{d}{4 f_0^2 r^{2\lambda}}
\sim \frac{l (l+1) + \gamma _{1,2}}{r^2},
\ee
should be nonsingular, i. e., $\gamma _{1,2}> - 1/4$; we consider
the core coupling
constants $\gamma _{1,2}$ as independent of $l$;\\
b) \quad the mapping of wave functions realized by  the operators $A^{\pm}$
of Eq.
(\ref{gena+})
\ba
\sqrt E \Psi_E^{(2)} = A^- \Psi_E^{(1)} &=&
\biggl( V_1 (r) - E + 2 f(r) \partial_r + 2 f'(r) + b(r) \biggr) \Psi_E^{(1)}
\nonumber\\
&=& \Bigl( 2 f^2 (r) - f'(r) - E + 2f(r) \partial_r \Bigr) \Psi_E^{(1)}
\ea
(where the basic Schr\"odinger equation has been applied)
should reproduce the correct physical behavior when
$r \rightarrow 0$ for a particular angular momentum $l$.

We obtain two possible behaviors for $f(r)$ in the vicinity of the
origin:\\
1) \quad  For $\lambda = - 1$  the consistency requirements
can be satisfied for all angular momenta $l_1$ only if $\gamma = 0$. The
related parameter $f_0$  takes the following values,
\ba
f_0 = - l_1 - 3/2; \quad \mbox{for} \quad l_2 = l_1 + 2; \\
f_0 = l_1 - 1/2; \quad \mbox{for} \quad
l_2 = l_1 - 2 \quad \mbox{if} \quad l_1 \geq 2.
\label{f01}
\ea
If $d < 0$ the HSSQM system can be embedded into a ladder of two standard
SSQM problems (see Eq. (\ref{ladder})) but the case  $d > 0$ is again
irreducible.\\
2) \quad For $\lambda = + 1$ the solution exists for $d < 0$ only which
can be always embedded into the ladder of standard SUSY systems with
raising and lowering of angular momentum so that as a result,
\be
l_1 = l_2 = l, \quad \mbox{and} \quad
f_0^2 = \frac{ - d}{(2l + 1)^2 + 4\gamma}; \quad \gamma _{1,2} = \gamma.
\label{f02}
\ee
For other values of $\lambda$ and $f_0$ either the operators $A^{\pm}$
map from the physical space into an unphysical one or a potential is
singular. In both cases the formal intertwining relations between
$h_1$ and $h_2$ do not result in the equivalence of energy spectra and
in the connection of physical wave functions.

The higher-order polynomial superalgebra for the radial problem can
be constructed in the same form as Eq.(\ref{exten}). However in this
case one obtains generally the spectral equivalence
between different partial waves when keeping in mind the preceeding
analysis.

\section{SSQM and scattering problem}
\subsection{One-dimensional scattering}
\hspace*{3ex}Consider the one dimensional scattering problem on
the line, i.e. $
x \in (- \infty, + \infty)$, for the Hamiltonian
\be
h_1 = - \partial^2 + V_1(x) = a^+ a^-
\ee
with a potential that reaches its constant asymptotic value (fast enough)
\be
V_1(x) \longrightarrow C \atop{x \rightarrow \pm \infty} \label{asymp}
\ee
where the bound state energy spectrum of $h_1$ is bounded by the
constant $C$ ($ 0 \le E_n \le C$)
\be
h_1 \Psi^{(1)}_n(x) = E_n \Psi^{(1)}_n(x); \hspace{1cm} \mbox{n= 0,1,..}
\ee
while the continuous spectrum is given by, $E(k) \ge C$,
\be
h_1 \Psi^{(1)}_k(x) = E(k) \Psi^{(1)}_k(x) ;
 \hspace{1cm} \mbox{$E(k) = k^2 + C .$}
\ee
The scattering wave function fulfils the asymptotic conditions
(see, for example \cite{chad})
\be
\Psi^{(1)}_{k, -\infty} = e^{+ikx} + R^{(1)}(k) e^{-ikx}
\ee
and
\be
\Psi^{(1)}_{k, +\infty} = T^{(1)}(k) e^{+ikx}
\ee
where $R^{(1)}(k),\,T^{(1)}(k)$ are the reflection and transmission
coef\/f\/icients respectively.

The ladder operators $a^{\pm}$ are asymptotically expressed as
\be
a^-_{\pm \infty} = - \partial + W_{\pm};\hspace{1cm}
a^+_{\pm \infty} = \partial + W_{\pm}
\ee
with
\be
W_{\pm} = \lim_{x \to \pm \infty} W(x) ;
\hspace{1cm} \mbox{$W^2_{\pm} = C$} \label{W-lim}
\ee
The asymptotic scattering wave function of the partner Hamiltonian
$h_2$ is then proportional~to
\be
a^-_{- \infty} \Psi^{(1)}_{k, -\infty} = (-ik + W_-) \
[e^{ikx} - R^{(1)}(k) \frac{k - iW_-}{k + i W_-} e^{-ikx}] \label{as-}
\ee
while
\be
a^-_{+ \infty} \Psi^{(1)}_{k, +\infty} = (-ik + W_-) \ [
T^{(1)}(k) \frac{k + iW_+}{k + i W_-} e^{ikx}] . \label{as+}
\ee
Hence the transmission and reflection coefficients associated to $h_2$
are respectively \cite{genref}, \cite{boy}
\be
T^{(2)}(k) = T^{(1)}(k) \frac{k + iW_+}{k + i W_-} \label{trans}
\ee
and
\be
R^{(2)}(k) = - R^{(1)}(k) \frac{k - iW_-}{k + i W_-} . \label{refl}
\ee
One can recognize, as is well known \cite{chad}, that the transmission
coefficient contains physical poles in the upper half of the complex
$k$-plane, their positions corresponding to the energies of bound
states $E_j = -\kappa^2_j + C, \, k_j = i\kappa_j$.
Thus the difference in physical pole structure of $T^{(1)}$ and
$T^{(2)}$ depends on the signs of $W_{\pm}$.
Unitarity always holds for $T^{(2)},\,R^{(2)}$ whenever it holds
for $T^{(1)},\,R^{(1)}$.

We now want to relate the above properties of transmission
coef\/f\/icient to the Witten index  introduced in the
previous section. Consider the equations
\be
a^{\mp} \Psi^{(1),(2)}_{E=0}(x) = 0
\ee
whose solutions are given in Eq.(8).
Three different cases arise
\begin{enumerate}
\item
$$W_- > 0 \hspace{.5cm} W_+ < 0 \hspace{.5cm} \Delta_W = 1 - 0 = 1
\hspace{.5cm} E_0^{(1)} = 0 \mbox{  and   } E_n^{(2)} > 0$$
\item
$$W_- < 0 \hspace{.3cm} W_+ > 0 \hspace{.3cm} \Delta_W = 0 - 1 = - 1
\hspace{.3cm} E_0^{(2)} = 0 \mbox{  and   } E_n^{(1)} > 0$$
\item
$$W_+ W_- > 0 \hspace{.5cm}   \Delta_W = 0  \hspace{.5cm}
E_n^{(1)} > 0 \mbox{  and   } E_n^{(2)} > 0$$
\end{enumerate}
 This establishes the following connection
\be
T^{(2)}(k = 0) = T^{(1)}(k = 0) \exp( i\pi \Delta_W)
\ee
between the relative phase of the supersymmetric partner amplitudes
and the Witten index (see also D.Boyanovsky and
R.Blankenbecler \cite{boy}).

The scattering problem for HSSQM, as defined in Eqs. (\ref{b2}), (\ref{V1,2})
 can be deduced from Eqs. (\ref{trans}),(\ref{refl}) by iteration:
\ba
T^{(2)}(k) = T^{(1)}(k) \frac{(k + iW_{2+})(k + iW_{1+})}
{(k + iW_{2-})(k + iW_{1-})}\nonumber\\
\nonumber\\
R^{(2)}(k) = R^{(1)}(k)\frac{(k - iW_{2-})(k - iW_{1-})}
{(k + iW_{2-})(k + iW_{1-})} \label{2scat}
\ea
where, assuming $f_{\pm} = \lim f(x)$ for $x \rightarrow \pm\infty$
are constant so that the potentials $V_{1,2}$ at $x \rightarrow
\pm\infty$ are finite  and following Eq. (\ref{W - f}), the asymptotic
values of the superpotentials are given by
\be
W_{1\pm} = \frac{c}{4f_{\pm}} - f_{\pm};
\hspace{1cm}W_{2\pm} = - \frac{c}{4f_{\pm}} - f_{\pm}.
\ee
In order to guarantee that the asymptotic values of the potentials $V_{i}$
both at $+ \infty$ and at $- \infty$ are equal, one must have
either $f^2_+ = f^2_-$ or $16 f^2_+ \,f^2_- = c^2$. The Witten index is
still determined by Eq. (\ref{deltaw}) and obviously can take
any integer value between 2 and - 1 for $c > 0$. As can be seen
from Eq. (\ref{fpm})
the breaking of the Witten criterion occurs if $f_+ = - f_-$ and
$16 f^4_{\pm} < c^2$.

The peculiar situation in which the Witten criterion is not valid appears
to be described by $c = 0$ and $W_{1\pm} = W_{2\mp} = 0 $.
\footnote{This case can provide examples of zero energy bound states.
There is the possibility that one partner has indeed a normalizable
state at zero energy whereas the companion state is not normalizable.
This is due to the fact that the relations between the norms involves
the square of the energy (i.e.0). We thus would realize an example of
partnership between a bound state and a continuum like state.} Although
the
wave functions, connected by the intertwining relations, are different
as well as the potentials, the spectrum and the phase shifts
of the two supersymmetric partners are identical.
This illustrates explicitely the fact that the knowledge of the phase shifts
and of the position of the bound states do not identify uniquely
a local potential.

For $d > 0$ and nonsingular potentials the function $f(x)$ is nodeless and
in order to have equal asymptotics Eq. (\ref{asymp}) one should impose
$f_+ = f_- = f_{\infty}$. Respectively the transmission and reflection
coefficients are connected as follows:
\be
T^{(2)} = T^{(1)} \hspace{1cm} \quad R^{(2)} = R^{(1)}\,\,
\frac{ (k + i  f_{\infty})^2 -
\frac{d}{4 f_{\infty}^2}}{(k - i  f_{\infty})^2 - \frac{d}{4f_{\infty}^2}}
\label{22scat}
\ee
The equality of the transmission coefficients is accounted for by
equal asymptotic values for each potential at $\pm\infty$,
as in (\ref{asymp}).

\subsection{Radial scattering problem}
\hspace*{3ex}The radial problem can be treated in a very close manner
except that the running variable $r$ varies from $0$ to $\infty$ and
the partner potentials $V_{1,2}$
go asymptotically to the constant value $C$. The value at the origin
may be finite or infinite but the $s$-wave wavefunction must go to $0$
linearly with $r$ as $r$ goes to $0$. Asymptotically one has then
\be
\Psi^{(1),l}_{k, \infty} = e^{-ikr} +  (-1)^{l + 1} S_l^{(1)}(k) e^{ikr}
\ee
and, in analogy with the preceding developments,
\be
a^-_{\infty} \Psi^{(1),l_1}_{k, \infty} = ( ik + W_{\infty}) \
[e^{-ikr} - (-1)^{l_1 + 1} S_{l_1}^{(1)}(k)
\frac{k + iW_{\infty}}{k - i W_{\infty}} e^{ikr}]
\ee
It is clear that the supersymmetric transformation introduces
singularities in the potential. Hence, there are two different interpretations
possible. One may consider a fixed $l$ partial wave, each supersymmetric
transformations adding a singular contribution, reminiscent of a centrifugal
barrier, to the preceding potential \cite{genref}. The second
interpretation is based on the recombination of the original potential with
the dynamically generated centrifugal barriers and hence the superpartner
is considered to be associated to a different orbital momentum $l$ (see,
for instance, F. Cooper et {\it al.} in ref. \cite{boy}). We will refer
from now on to this last interpretation.

If $W_{\infty} \neq 0$ the
$S-$matrix element $S_{l_2}^{(2)}(k)$ associated to the partner
Hamiltonian is given~by
\be
S_{l_2}^{(2)}(k) = -  S_{l_1}^{(1)}(k) \frac{k + iW_{\infty}}{k -
i W_{\infty}}, \qquad l_2 = l_1 .
\label{smat}
\ee
reflecting the presence
\footnote{For a discussion see Amado et al.Int.J.Mod.Phys.
quoted
in Ref.8 }.
of long-range forces $\sim r^{-3}$
In this case the behaviour of $ S_{l_2}^{(2)}(k)$
for small $k$ is of $s-$wave type independently of the angular
momentum.
We remind that the correct behaviour of wave functions
at the origin requires that $l_2 = l_1 \pm 1$. It can be made compatible
with Eq.(\ref{smat}) only at the expense of introducing a
centrifugal core which depends on angular momentum.

On the other side for $W_{\infty} = 0$ the two $S$
-matrices coincide
\be
S_{l_2}^{(2)}(k) =  S_{l_1}^{(1)}(k),
\qquad l_2 = l_1 \pm 1.
\label{smatt}
\ee
signaling an anomalous threshold behaviour related
to the existence of zero energy pole which makes the effective range
expansion invalid \footnote{for a discussion see Amado et al.Phys.Rev.
C43 quoted in Ref.8}.

The radial scattering problem for the second order derivative SUSY system
can be readily solved if $d = 0$,
by repetition of the single SSQM relations.
Evidently $S$-matrices will coincide belonging to
partial waves $l_2 = l_1, l_1 \pm 2$.

Let us analyze the features of scattering problem for HSSQM Hamiltonians
when $d \not= 0$. We assume the following asymptotics for $f(r)$:
\be
f(r) = \bar f_0 + \bar f_1 r^{\lambda} + \bar f_2 r^{2\lambda} + \cdots;
\quad r \rightarrow \infty,
\ee
where $\bar f_{0,1} \not= 0$. Let us restrict ourselves to
potentials decreasing at least as fast as its centrifugal part (or faster
in the $s$-wave) that gives the bound  $\lambda \leq - 1$.

We have found four essentially different types of potentials depending
on their asymptotics.\\
a) The case $\lambda = -1$  is consistent for
\be
d = - 4 \bar f_0^4 < 0, \label{dfbar}
\ee
if the Coulomb-type long range forces are absent.
The relevant SUSY transformation
gives rise to the following change in the angular momentum,
\ba
l_2 &=& l_1 - 1, \quad \bar f_1 = \frac{1}{2} l_1;\nonumber\\
l_2 &=& l_1 + 1, \quad \bar f_1 = - \frac{1}{2} (l_1 + 1).
\label{l1a}
\ea
When comparing with the requirements on the wave function at the origin,
Eqs. (\ref{f01}), (\ref{f02}), one concludes that they can be satisfied if
the related potentials have short range core of centrifugal type which
depends on angular momentum.\\
b) In the  case $- 2 < \lambda < -1$ the consistency is again assured by
Eq. (\ref{dfbar}) and the potentials fall off $ \sim r^{\lambda -1}$ and
display $s$-wave behavior.\\
c) For $\lambda = -2$, whatever is the sign of $d$, the centrifugal term is
not changed under the  HSUSY transformation and the coefficients $\bar f_i$
obey the relation,
\be
2 \bar f_0 \cdot \bar f_1 \Bigl( 1 + \frac{d}{4\bar f_0^4}\Bigr) = l (l+1).
\ee
d) For $\lambda < -2$, only $s$-wave is allowed.

Thus the consistent treatment of wave functions behaviour at the
origin and at the infinity selects out the only possibility
$l_1 = l_2, d < 0$. The irreducible transformation $d > 0$
induces the short range forces depending on angular momentum.
The connection between $S$-matrices of superpartner systems is
given by,
\be
S_{l_2}^{(2)} = S_{l_1}^{(1)} \frac{k^2 - 2ik \bar f_0 - \bar f_0^2 -
\frac{d}{4\bar f_0^2}}{k^2 + 2ik \bar f_0 - \bar f_0^2 -
\frac{d}{4\bar f_0^2}},
\ee
and, generally, they differ by two poles.

The potentials belonging to
the "a" or "b"-type   only differ by one pole because  of the constraint
(\ref{dfbar}), so that
\be
S_{l_2}^{(2)} = S_{l_1}^{(1)} \frac{k - 2i \bar f_0}{k + 2i \bar f_0},
\ee

\section{$q$-deformed SUSY quantum mechanics}
The concept of $q$-deformations has drawn recently much attention
as a way to extend the description of symmetries based on Lie algebras
\cite{rtf}. In application to SSQM, an extension based on q-deformed
SUSY algebra has been developed by V. Spiridonov \cite{spir} in a
particular realization exploiting the dilatation of coordinates.

The main tool  for a $q$-deformation is given by the dilatation operator,
\be
T_q f(x) = \sqrt q f(qx),\quad T_q \partial_x = q^{-1} \partial_x T_q,\quad
\ee
which can be represented by the following pseudodif\/ferential operator
\be
T_q = \sqrt q \exp (\ln q\enspace x\partial_x),
\quad T_q^{\dagger} = T_q^{-1}.
\ee
 Let us now introduce the $q$-deformed components of supercharges $Q^{\pm}$,
\be
a_q^+ =(\partial + W(x))\,T_q,\quad a_q^- = T_q^{\dagger}
(- \partial + W(x)), \label{aqpm}
\ee
in the notations of Section 2. The Hamiltonian components (see Eq.
(\ref{h1h2})) become now the $q$-deformed
\ba
h_1 & = &  a_q^+ a_q^- = - \partial^2 + W^2 (x) + W'(x),\nonumber\\
h_2 & = & q^{-2} a_q^- a_q^+  = - \partial^2 + q^{-2} W^2 (q^{-1}x)
- q^{-1} W'_x (q^{-1}x),   \label{qH}
\ea
if the kinetic terms are properly normalized. Evidently these Hamiltonians
are not intertwined by means of the standard SUSY algebra (see Eq.
(\ref{intertw})).
Rather they obey the $q$-deformed intertwining relations,
\be
h_1 a_q^+ = q^2 a_q^+ h_2,\quad a_q^- h_1 = q^2 h_2 a_q^-.
\ee
The complete $q$-deformed SUSY algebra has the conventional form, Eq.
(\ref{basic1})
but with the $q$-(anti)commutators instead of the usual ones,
\be
[ X, Y]_q \equiv  X Y - q^{-2} Y X,\quad \{X, Y\}_q \equiv
 X Y + q^{-2} Y X.
\ee
In particular, one has
\be
\{ Q^+, Q^-\}_q = H;\quad [Q^+, H]_q = [H, Q^-]_q = 0. \label{qalg}
\ee

Obviously the supercharges are not conserved now in the usual sense because
they do not commute with the Hamiltonian. As a consequence the superpartner
Hamiltonians are no more isospectral but their spectra are
related by  $q$-dilatation, $E^{(1)} = q^2 E^{(2)}$. We notice
however that there exists a
$q$-Hamiltonian, $H_q$ , which  commutes with supercharges and obeys the
superalgebra,
\be
\{ Q^+, Q^-\} = H_q \equiv
 \left(\begin{array}{cc}
h_{1} & 0 \\ 0 & q^2 h_{2} \end{array}\right);\quad
[H_q, Q^{\pm}] = 0. \label{qham}
\ee

The wave functions are unambiguously interrelated by the action of the
operators $a_q^{\pm}$,
\be
\psi^{(2)} \propto a_q^- \psi^{(1)},\quad \psi^{(1)} \propto a_q^+ \psi^{(2)}
\ee
Obviously the Witten-type analysis of spontaneous SUSY breaking can be
extended for $q$-deformed SUSY, in terms of $q$-Hamiltonian
$H_q$, in full analogy with the standard SSQM in Section 2.

The above realization of $q$-deformed SUSY is flexible
enough  to be easily combined with other extensions of SSQM such as
the paraSUSY (\cite{pss}) and HSUSY (see Section 3) quantum mechanics.
The particular (factorizable) construction of
the second order derivative $q$-deformed SUSY algebra can be
realized by means of a sequence of two $q$-deformations (\ref{aqpm})
with different dilatation parameters $q_1, q_2$ . The relevant components of
the $q$-supercharge are given by products,
\be
A^+_q = q_1 a^+_{q_1} \cdot a^+_{q_2}
= q_1 (\partial + W_1(x))\,T_{q_1} (\partial + \widetilde W_2(x))\,T_{q_2} =
( A^-_q)^{\dagger}.
\ee
The supercharge components can be transformed to the form of
second order derivative operator augmented with the single dilatation,
\ba
A^+_q &=& (\partial + W_1(x)) (\partial + W_2(x))\,T_{q};\nonumber\\
q &=& q_1 \cdot q_2;\quad W_2 (x) \equiv q_1 \widetilde W_2 (q_1 x).
\ea
In such a form the generalization of
second order derivative SUSY is straightforward and
can be performed following the scheme of Section 3 with the dilatation at the
last step both for the reducible ($d < 0$) and for the irreducible
($d > 0$) cases.

As in the ordinary $q$-deformed case the Hamiltonian $H$ does not
commute with the supercharge but it satisfies Eqs. (\ref{qalg}).
The $q$-deformed SUSY algebra with the
 Hamiltonian $H$ reads

$$Q^+ Q^- + q^{-4} Q^- Q^+ = \{Q^+, Q^-\}_{q^2} = (H -
\alpha q^{\sigma_3 - 1})^2  + d \cdot q^{2(\sigma_3 - 1)},$$
\be
[Q^+ , H]_q = [ H, Q^- ]_q = 0,
\label{qhsusy}
\ee
where $\sigma_3$ is the Pauli matrix.

But again there
exists a $q$-Hamiltonian $H_q$, Eq. (\ref{qham}),
which  commutes with the supercharges. Moreover in terms of $H_q$ the
HSUSY algebra takes the usual form, Eq. (\ref{b2}).

Further steps in extension
of the higher order derivative SUSY either
 lead to Eq. (\ref{exten}) with the $q$-Hamiltonian and the
 conventional SUSY or to its $q$-deformed
 version \cite{ais} with the true Hamiltonian $H$ and
 the primitive blocks defined in (\ref{qhsusy}),

$$ \{Q^+, Q^-\}_{q^n} = \prod_{i + 2j = n}  ( H - c_i \cdot
 q^{\sigma_3 - 1})
 \Bigl((H - \alpha_j q^{\sigma_3 - 1})^2
 + d_j \cdot q^{2(\sigma_3 - 1)}\Bigr);$$
 \be
 [ Q^+ , H ]_q = [ H, Q^- ]_q = 0;
 \quad d_j > 0.
\ee

Concerning the scattering problem for the above models,
the asymptotic relations, Eqs.
(\ref{as-}), (\ref{as+}), (\ref{trans}), (\ref{refl}) should
be modified  taking into account Eq. (\ref{aqpm}). As a
result one finds the connection between transmission and reflection
coef\/ficients of $h_1$ and $h_2$ which now involves the dilatation
parameter $q$,
\ba
T^{(2)}(k) = T^{(1)}(k\,q) \frac{(k
q + iW_{+})} {(kq + iW_{-})}\nonumber\\
R^{(2)}(k) = -
R^{(1)}(k\,q)\frac{(k q - iW_{-})} {(k q + iW_{-})}  \label{qscat}
\ea
Thus both in the
discrete and in the continuous parts of spectrum we remove the spectrum
degeneracy typical for the ordinary SSQM and come to scaling relations.

In the framework of $q$-deformed SUSY mechanics the
problem of the construction of self-similar potentials has recently
been tackled \cite{spir} and
a class of such potentials has been found \cite{spir}, \cite{shab}.
They are obtained from solutions of ladder equations with $q$-periodic closure
conditions. For instance, in the case of the ordinary $q$-deformed SUSY
mechanics (\ref{qH}) they lead to the following equation for the
superpotential \cite{shab},
\be
V_1 (x) = W' (x) + W^2 (x) = V_2 (x) + c = - q^{-1} W' (q^{-1} x) +
q^{-2} W^2 (q^{-1} x) + c.
\ee
In the scattering regime the potentials are decreasing at the infinity and
therefore $c = W_{\pm}^2 (1 - q^{-2})$.

The $q$-deformation of scattering data allows us to formulate the dual
self-similarity condition which in general does not imply the self-similarity
of potentials. We formulate this condition again as a closure
condition like \cite{shab} but for scattering amplitudes in
momentum space,
\be
T^{(2)} (k) = T^{(1)} (k)  ;\quad
R^{(2)} (k) =  R^{(1)} (k),
\ee
which is combined with Eq. (\ref{qscat}). Their solutions for $  q > 1$
and $W_+ = - W_- > 0$ are
\ba
T^{(1)}_{q>1}(k) \equiv T(q, \frac{W_+}{k}) &=& \prod_{n =1}^{\infty}
\frac{(k + i \frac{W_+}{q^n})}
{(k - i \frac{W_+}{q^n})} t(\ln k)\equiv {}_1\Phi_0 (- 1; i W_+/k)
t(\ln k) ;
\nonumber\\
R^{(1)}_{q>1}(k) \equiv R(q, \frac{W_+}{k}) &=& \prod_{n =1}^{\infty}
\frac{(k + i \frac{W_+}{q^n})}
{(k - i \frac{W_+}{q^n})} r(\ln k)\nonumber\\
 \equiv {}_1\Phi_0 (- 1; i W_+/k)
 r(\ln k), \label{RT}
\ea
where $ |t|^2 + |r|^2 =1$ and $t(z) ( r(z) )$ are periodic (antiperiodic)
functions,
$$ t (z + \ln q) = t (z), \quad  r (z + \ln q) = - r (z)$$
and the self-similar transmission and reflection coefficients
are parametrized by
the hypergeometric function, ${}_1\Phi_0 ( a; z )$ \cite{bat}.
If $0 < q < 1$ and $W_+ = - W_- < 0$ then the consistent solution is
expressed in terms of the previous one, Eqs.(\ref{RT}),
\be
T^{(1)}_{q<1}(k) = T(1/q, -\frac{W_+}{k})\frac{k - iW_+}{k + iW_+}
\nonumber
\ee
and a similar equation takes place for $R^{(1)}_{q<1}(k)$.

If one imposes the condition of vanishing of the reflection coefficient
 for high $k$ ,which amounts to the validity of the Born expansion,
\cite{chad}, one is lead to  assume $r(\ln k) =0$ and
to put  $t(\ln k)=1$, respectively.
It is thus interesting to analyze the interconnections between
self-similarities of reflectionless potentials and of scattering amplitudes
 \cite{barclay}. It is clear that the underlying reflectionless potential
has infinitely many bound states with an accumulation point around zero.
As the Born expansion is valid it means that such a potential is slowly
decreasing and oscillating.
Our construction is connected to recent
investigations concerning a class of reflectionless potentials with infinitely
many bound states \cite{barclay}, we stress however that our
starting point has been the exploration of the condition
 of self-similarity for the scattering coefficients
leaving aside the detailed $x$ dependence  of the
(super)potentials~\footnote{see also \cite{spir} for an exponential like
spectru
having the same $T$-matrix as we have.}~.

One can easily derive $q$-deformed HSUSY relations between
the transmission and reflection coefficients
guided by similar arguments as those leading to  (\ref{2scat}) and
 (\ref{22scat}).

 In conclusion we remark that both the polynomial SSQM and
 $q-$deformed SSQM may be generalized onto matrix potentials
 and partially extended onto higher space dimensions. Some of
 extensions are considered in \cite{david} and $q-$deformations
 of polynomial SSQM will be described elsewhere.

A.A.A., J.-P.D. and M.V.I. are indebted to the University of
Bologna and INFN for the support and hospitality. Two of us
(A.A.A. and M.V.I.) were supported by the Russian Foundation
for Fundamental Research (Grant No.94-01-01157-a).

\end{document}